\documentstyle[aps,multicol,epsf]{revtex}
\catcode`\@=11

\def\eqbegin         {  \begin{eqnarray}  }
\def\eqend           {  \end{eqnarray}  }
\def\beq{\begin{equation}}
\def\eeq{\end{equation}}

\def\del          { \partial }

\def\Z{{\bf Z}}

\def\hs_2{\hspace{2mm}}
\def\hs_3{\hspace{3mm}}
\def\hsf{\hspace{4mm}}

\title{Pairing Effects in the  Edge  of   Paired Quantum Hall States}
\author{Kazusumi Ino \thanks{e-mail:ino@kodama.issp.u-tokyo.ac.jp}}

\begin{document}
\maketitle
\begin{center}
{\it Institute for Solid State Physics, University of Tokyo,} \\
{\it Roppongi 7-22-1,  Minatoku,  Tokyo,  106, Japan} 
\end{center}
\thispagestyle{empty}
\begin{abstract}
We study pairing effects in the edge states of  paired 
fractional quantum Hall states by using persistent edge currents 
as a probe.   We give the grand partition functions for edge
excitations of paired states
 (Haldane-Rezayi, Pfaffian, 331) coupling to an  Aharanov-Bohm flux and 
 derive the exact formulas of the persistent edge current.   
We show that the currents are flux periodic with 
the unit flux $\phi_0=hc/e$. 
At low temperatures,  they exhibit 
anomalous  oscillations in their flux dependence. 
The shapes of the functions depend on the bulk topological 
order.  They  converge to the sawtooth 
function with   period $\phi_0/2$ at 
zero temperature, which indicates pair condensation. 
This phenomenon provides  an interesting bridge between  
superconductivity in 2+1 dimensions and superconductivity 
in 1+1 dimensions.    
We propose experiments of measuring the persistent current   
 at even denominator plateau  in single or 
 double layer systems to test our predictions.   
\end{abstract}

\begin{multicols}{2}
One of the surprising aspects of the fractional quantum Hall 
effect is that the edge state forms  a new kind of  state  of matter 
beyond Fermi liquid, called the  chiral Tomonaga-Luttinger liquid \cite{wen}.  
 Some experiments have already  demonstrated  the characteristic behavior of 
chiral Tomonaga-Luttinger liquid \cite{kane,milli}.   
Recently the Aharanov-Bohm effect (AB effect) in such  systems  were 
studied  by Geller et al  
\cite{geller,geller2} and Chamon et al \cite{chamon}.  
Especially, the former authors predict new edge-current
 oscillations in the persistent current at the edge of 
 the $\nu=\frac{1}{q}$ Laughlin state,   
 which has   
no amplitude reduction from disorder and thus results in 
a universal non-Fermi-liquid temperature dependence \cite{geller2}. 
Also the persistent current for the annulus Laughlin state was 
recently investigated by Kettemann \cite{ketteman}. These 
studies show the current is periodic with a unit flux quanta
$\phi_0=hc/e$ in agreement with the theorem of Byers and Young
\cite{byer}. 

Motivated by these recent studies, 
 we  investigate  the persistent currents 
in paired fractional Hall states in this paper.  
 The states we will consider 
are the 331 state \cite{halp}, the Haldane-Rezayi state \cite{haldane2} 
 and the Pfaffian state \cite{moore}.  They are 
quantum Hall analogs of the BCS superconductor. The pairing 
symmetry is p-wave with $S_z=1,0$ for  the Pfaffian and the 331 states 
respectively, and 
d-wave for  the Haldane-Rezayi states.    
 The Haldane-Rezayi state and the Pfaffian state are 
 new kinds of quantum Hall state which recently 
 attracted considerable attentions because 
 they are supposed to exhibit 
 some novel features beyond ordinary quantum Hall state 
 in their topological ordering,  such as nonabelian statistics and 
 specific degeneracy on a surface with nontrivial topology.    
 On the other hand, the 331 state is a part of generalized hierarchy 
\cite{wenzee}, but 
 can be interpreted as a paired state. 
  These states are proposed as 
a candidate for the $\nu=\frac{5}{2}$  plateau 
 in single layer systems \cite{willet} 
and  the $\nu=\frac{1}{2}$ plateau in double 
layer systems \cite{he}. Recent numerical study by Morf suggests 
that the $\nu=\frac{5}{2}$ state is Pfaffian-like \cite{morf}.

As in the Laughlin state, these states are incompressible and 
have edge excitations. 
The edge states have  a richer content than 
chiral Tomonaga-Luttinger liquid  due to their internal degree of freedom 
\cite{wen2,wenwu,milo,gurarie,lee}.  
Since one cannot see into the bulk state directly,    
it is important to know how 
the edge state reflects the bulk topological order.    
Thus we will compare the persistent edge currents for paired states
to see how the pairing of the electrons in the bulk 
affects the properties of the edge states. 
Since   bulk states of paired states are supposed to be 
in a  superconducting phase, 
one naively imagine that the persistent edge 
current should have a period $\phi_0/2
=hc/2e$.  However the fact that 
  there is no spontaneous
symmetry breaking of continuous symmetry in $1+1$ dimensional 
quantum field theory prevent this naive guess to be right. 
We will examine the property of pairing in the edge state 
 by using the edge conformal field theory.

Let us first recall edge excitations on a single edge of    
the $\nu=\frac{1}{q}$ Laughlin state.    
The edge state of the Laughlin state is a 
chiral Tomonaga-Luttinger liquid which is described by a chiral boson $\varphi$.  
Its Hilbert space is described by 
 the $U(1)$ Kac-Moody algebra generated by $j=\frac{1}{\sqrt{q}}\del \varphi$ 
 and the zero modes which correspond to quasiparticles. 
The Hamiltonian is given by  
$H=\frac{1}{2}\sum_{n \in Z} j_{n}j_{-n}-\frac{c}{24}$  
where we include a Casimir factor with $c=1$. 
The complete description of the edge 
excitations can be given  by  a  Virasoro character which 
corresponds to the partition function of the grand ensemble of 
quasiparticles.  For the edge state of the Laughlin state, 
the Virasoro character is given by  
\eqbegin
\chi_{q}(\tau) = \frac{1}{\eta}\sum_{n \in \Z}
{\rm exp}\left( \frac{\pi i}{q} \tau  n^{2}\right) 
\label{ZL}  
\eqend 
where $\eta$ is the Dedekind function 
$\eta(\tau)=x^{\frac{1}{24}}\prod_{n=1}^{\infty}(1-x^n),\hsf  
x=e^{2\pi i \tau}$ and $\tau$ is a modular parameter. 
Let us modify (\ref{ZL}) to get 
the grand partition function for the edge state 
coupling to  an AB flux with finite size effects. 
Let $L$ be the circumference of the edge state. 
The finite size $L$ induces
 a  temperature scale $T_0=\frac{\hbar v}{ k_{B}L}$ 
where $v$ is the  Fermi velocity of the edge modes 
determined by the confining potential.   
For example, a Fermi velocity of $10^{6}$ cm/s and circumference of
$1\mu $ m yields $T_0 \sim  60$ mK.
The coupling of  the edge state to an AB flux $\Phi$ 
 induces a twisted boundary condition on the chiral 
boson $\varphi$. 
Accordingly   the grand partition function 
 becomes \cite{cappelli,geller2}     
\eqbegin 
Z_{\rm Laugh}(\tau,\phi) = \chi_q(\tau,\phi)&=&\frac{1}{\eta} 
\sum_{n \in \Z}{\rm exp}\left( \frac{\pi i}{q}\tau(n-\phi)^{2}\right)
\nonumber\\ 
&=&\frac{\sqrt{q}}{\eta(t)}\theta_3(\phi |qt)
\label{Zs}
\eqend 
where  $\tau=i\frac{T_0}{T}, t=i\frac{T}{T_0}$, 
$\phi=\Phi/\phi_0$ with  $\phi_0=hc/e$  the unit 
flux quantum and $\theta_3 $ is the third Jacobi $\theta $ function.
From the grand partition function Eq.(\ref{Zs}), 
one can deduce the exact formula of the persistent current   
for a chiral Tomonaga-Luttinger liquid. 
In general,  
the persistent current $I$ is defined by the following formula:  
\eqbegin 
I\equiv \frac{T}{\phi_0}
\frac{\del {\rm ln}Z(\tau,\phi)}{\del{\phi}}. 
\label{perfor}
\eqend 
Then the persistent current in the Laughlin state is calculated to be   
\eqbegin 
I_{\rm Laugh} =\frac{2\pi T}{\phi_0}\sum_{n=0}^{\infty} (-1)^{n}  
\frac{\sin(2\pi n\phi)}{\sinh(nq\pi T/T_0)} 
\label{pers}
\eqend 
which is the formula obtained \cite{geller2}.  
As the temperature is lowered, 
 the shape as a function 
of $\phi$ changes from a sinusoidal to 
the sawtooth function, 
\eqbegin 
I_{\rm Laugh} &\rightarrow& \frac{T_0}{q\phi_0}\sum \frac{(-1)^{m}}{m}{\rm sin}2\pi
m\phi \\ &=& 
-\nu \frac{ev}{L}(\phi-r),
\label{Lsaw}
\eqend 
for   $-\frac{1}{2}+r<\phi< \frac{1}{2}+r, \hsf r \in \Z.  $
The periodicity of $I_{\rm Laugh}$ in $\Phi$ is $\phi_0$, which agrees with 
the general theorem of Byers and Yang \cite{byer}. This 
is due to the presence of the quasiparticle with 
a fractional charge as argued in Refs.\cite{geller2,ketteman}. 
Since there is no backscattering from impurities
 in the chiral luttinger liquid, the current has no reduction from 
impurities and therefore shows non-Fermi liquid dependence
 on the temperature.   

Let us now consider paired quantum Hall states.
We first deduce the grand partition function for 
the edge excitations of a single edge of  the paired states coupling 
to an AB flux with the finite size effect.  

In general, paired states have an extra internal degree of freedom other than 
the charge. In the bulk conformal field theory description, 
paired states have some kinds of fermion $\psi$ for 
an internal degree of freedom  and a chiral boson for 
the charge degree of freedom.  The operator for the particles of the 
system is  of the  form $\psi e^{i\sqrt{q}\varphi}$ with 
$q$ is even for the system of electrons.  The edge excitations 
accordingly have a neutral sector corresponding to $\psi$.

 Consider  the Pfaffian  state.   
The additional sector of edge excitations of this state is 
described by Majorana-Weyl fermion. Accordingly,  it  
is generated by the $c=1/2$ Virasoro algebra. 
Since there are degenerate states in its spectrum, 
these  modes contribute to  the partition function
 through the following Virasoro characters for  Majorana-Weyl
 fermion: 
\eqbegin 
\chi^{\rm MW}_1(\tau)&=&\frac{1}{2}x^{-\frac{1}{48}}
\left(\prod_{0}^{\infty}(1+x^{n+\frac{1}{2}}) +
 \prod_{0}^{\infty}(1-x^{n+\frac{1}{2}}) 
 \right), \\ 
 \chi^{\rm MW}_{\psi}(\tau)&=&
 \frac{1}{2}x^{-\frac{1}{48}}\left(\prod_{0}^{\infty}(1+x^{n+\frac{1}{2}}) -
 \prod_{0}^{\infty}(1-x^{n+\frac{1}{2}}) 
 \right),\\ 
 \chi^{\rm MW}_{\sigma}(\tau)&=&
 x^{\frac{1}{24}}\prod_{1}^{\infty}(1+x^n).
\eqend 
$\chi^{\rm MW}_{1}$  and $\chi^{\rm MW}_{\psi}$ correspond to 
the untwisted sector of Majorana-Weyl fermion and 
$\chi^{\rm MW}_{\sigma}$  to the  twisted sector.  
The couplings of these sectors 
to the charge are determined by the condition of no monodromy with 
$\psi e^{i\sqrt{q}\varphi}$,  
which requires the untwisted sector couple to $\Z/q$  charge modes 
 and twisted sector to $(\Z+1/2)/q$ charge modes. 
Therefore the couplings are given by following 
products of  characters 
\eqbegin 
\chi_{1}^{\rm MW}\chi_{q}, \hsf \chi_{\psi}^{\rm MW}\chi_{q}, \hsf  
\chi^{\rm MW}_{\sigma}\chi^{(1/2)}_{q}, 
\eqend 
where we have introduced a  character
\eqbegin 
\chi^{(1/2)}_q(\tau)=\frac{1}{\eta}\sum_{n\in \Z+1/2}
 {\rm exp}\left( \frac{\pi i} {q} \tau n^2\right). 
\eqend    
Now the character of edge excitations of the Pfaffian state becomes  
\eqbegin             
\chi_{\rm Pf} &=& \chi_1^{\rm MW}\chi_{q}+\chi_{\psi}^{\rm
  MW}\chi_{q}+\chi_{\sigma}^{\rm  MW}\chi^{(1/2)}_{q}.
\label{Zpf}
\eqend 
To deduce the grand partition functions with 
 an AB flux, we note that 
the AB flux will couple only to the charged sector of 
the edge states and change the boundary condition of 
the chiral boson $\varphi$ as in the Laughlin state,  
which changes the grand partition function of chiral boson 
 to  $\chi_{q}(\tau,\phi)$ of Eq.(\ref{Zs}). 
 Also  $\chi^{(1/2)}_{q}(\tau)$ 
changes to  
\eqbegin 
\chi^{(1/2)}_{q}(\tau,\phi)&=&\frac{1}{\eta}\sum_{n\in \Z+1/2}
 {\rm exp}\left( \frac{\pi i} {q} \tau (n-\phi)^2\right) \nonumber\\
&=&\frac{\sqrt{q}}{\eta(t)}\theta_4(\phi|qt), 
\eqend 
where $t=i\frac{T}{T_0}$.   Then the grand partition functions 
with an AB flux  can be written in terms of 
Jacobi $\theta$ functions as 
\eqbegin 
Z_{\rm Pf}(t,\phi)&=&\frac{\sqrt{q}}{\eta^{3/2}}
\left[\sqrt{\theta_{3}(0 |t)}\theta_{3}(\phi |qt)
+\sqrt{\frac{\theta_{2}(0 | t)}{2}}\theta_{4}(\phi | qt)\right] 
\nonumber\\ 
\label{ZpfJ2} 
\eqend 
Similarly the Haldane-Rezayi state and the 331 state have 
the internal degree of freedom described by  
the $c=-2$ scalar fermion and the $c=1$ Dirac fermion respectively 
\cite{milo}.   The grand partition functions become 
\eqbegin
Z_{\rm HR}(t,\phi)&=& \frac{\sqrt{q}}{\eta^2}\left[\theta_{4}(0 |
  t)\theta_{3}(\phi |qt)+\theta_{3}(0 |t)
\theta_{4}(\phi | qt) \right],  
\label{ZhrJ2}
\\
Z_{331}(t,\phi)&=&\frac{\sqrt{q}}{\eta^2}
\left[\theta_{3}(0 | t) \theta_{3}(\phi |qt) +
\theta_{4}(0 |t)\theta_{4}(\phi | qt)\right].  
\label{Z331J2}
\eqend 
Here the zero modes of fermions are included.  
We see that $Z_{\rm HR}(\tau,\phi\pm 1/2)=Z_{331}(\tau,\phi)$. 
This implies that the the 331 state continuously evolve into 
the Haldane-Rezayi state by activating an AB flux by half unit 
flux quantum.

Putting these formulas into  Eq.(\ref{perfor}), we get  
the exact formulas for the persistent current for the 
Pfaffian, Haldane-Rezayi, 331 states as   
\eqbegin  
I_{\rm Pf}(t,\phi)&=&
\frac{T}{\phi_0}\frac{\sqrt{\theta_{3}(0 |t)}
\theta'_{3}(\phi |qt)
+\sqrt{\frac{\theta_{4}(0 | t)}{2}}\theta'_{4}(\phi | qt)}
{\sqrt{\theta_{3}(0 | t)}\theta_{3}(\phi |qt)
+\sqrt{\frac{\theta_{4}(0 | t)}{2}}\theta_{4}(\phi | qt)}. 
\label{ipf2}
\\
I_{\rm HR}(t,\phi)&=&\frac{T}{\phi_0}\frac{\theta_{4}(0 |
  t)\theta'_{3}(\phi |qt)
+\theta_{3}(0 | t)
\theta'_{4}(\phi | qt)}{\theta_{4}(0|t)\theta_{3}(\phi |qt)
+\theta_{3}(0 | t)
\theta_{4}(\phi | qt)}, \label{ihr2}\\ 
I_{\rm 331}(t,\phi)&=&\frac{T}{\phi_0}\frac{\theta_{3}(0 |
  t) \theta'_{3}(\phi |qt) 
+\theta_{4}(0 |t)\theta'_{4}(\phi | qt)}
{\theta_{3}(0 |t) \theta_{3}(\phi |qt) 
+\theta_{4}(0 |t)\theta_{4}(\phi | t)}.
\label{i3312}
\eqend 
Here $\theta'(\phi|\tau)$ is the differential of $\theta(\phi|\tau)$
with respect to $\phi$. 
We see $I_{\rm HR}(t,\phi \pm 1/2)=I_{331}(t,\phi)$.

The analytic properties of these currents follows from the 
properties of $\theta$ functions.   
At low temperature $T\ll T_0$, the contribution from $\theta_3$ 
which can be  considered as from the chiral Tomonaga-Luttinger
liquid is dominant in $I$ around $\phi \sim  0$ and its 
periodic points.   On the other hand, 
around $\phi \sim 0.5$, $\theta_4$ which is from the pairing 
structure is dominant.  This is because 
 $\theta_3$ ($\theta_4$) is localized around points $\phi \in \Z$
($\phi \in \Z+1/2$) as the temperature is 
lowered.   
 Therefore the currents converge at zero temperature to 
\eqbegin 
I &\rightarrow&  
-\nu \frac{ev}{k_BL}(\phi-\frac{1}{2}r),  
\label{Psaw}
\eqend  
for $-\frac{1}{4}+\frac{1}{2}r<\phi< \frac{1}{4}+\frac{1}{2}r, 
\hspace{3mm} r \in \Z$.    
At high temperatures, $\theta_3$ and $\theta_4$ are not 
localized and thereby their contributions are tamed. Thus 
the currents are suppressed compared to 
the current for  the Laughlin state.

\begin{figure}
\noindent
\hspace{1.5 in}
\epsfxsize=3.5in
\epsfbox{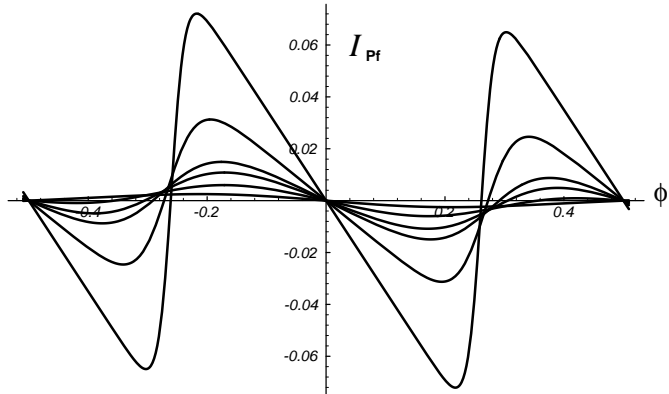}
\caption[flux1_2]
{The flux dependence of the  persistent currents for the $\nu=1/2$ Pfaffian state 
at temperatures $T/T_0=0.6,0.5,0.45,0.45,0.37,0.3$. 
The currents are measured in the unit $\nu ev/2(k_BL)$}
\vspace{.2cm}
\label{pf2}
\end{figure}

We especially take the simplest and experimentally important $\nu=1/2$
case to show some plots. Similar behavior holds for arbitrary $q$.  
Fig.\ref{pf2} 
 shows the flux  oscillations of 
persistent currents $I_{\rm Pf}$ 
for the $\nu=1/2$ Pfaffian state at 
$T/T_0=0.6,0.5,0.45,0.37,0.3$.  
The current  is   periodic with period $\phi_0$, but 
the flux dependence exhibits an anomalous behavior in  
these cases.  
As the temperature is lowered, extra zero points appear and 
the shape as a function of $\phi$ is no longer a sinusoidal function. 
These extra zero points approach to the points $\pm \phi_0/2$ as 
the temperature is lowered.  
These points equal to  $\pm \phi_0/2$  only at 
 zero temperature and the shape as a function of 
$\phi$ converges to the sawtooth function in (\ref{Psaw}) 
which have a period $\phi_0/2$, which means that 
a pairing condensation occurs. 

The currents for the Haldane-Rezayi state and the 331 state 
exhibit similar behaviors.  

\begin{figure}
\noindent
\hspace{1.5 in}
\epsfxsize=3.5in
\epsfbox{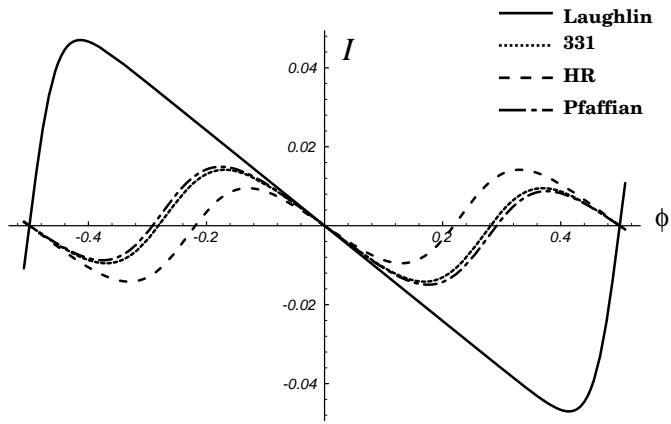}
\caption[teperaturenu1_2]
{The oscillations of persistent currents  at $T/T_0=0.45$.}
\label{p3h}
\end{figure}

The currents at $T/T_0=0.45$ are compared in Fig.{\ref{p3h}}. 
It shows that the shape of the oscillations
 is different for each state at finite temperatures 
and  therefore depend on the bulk topological order.    
We also see that the currents are suppressed compared to 
the Laughlin state. 

\begin{figure}
\noindent
\hspace{1.5 in}
\epsfxsize=3.5in
\epsfbox{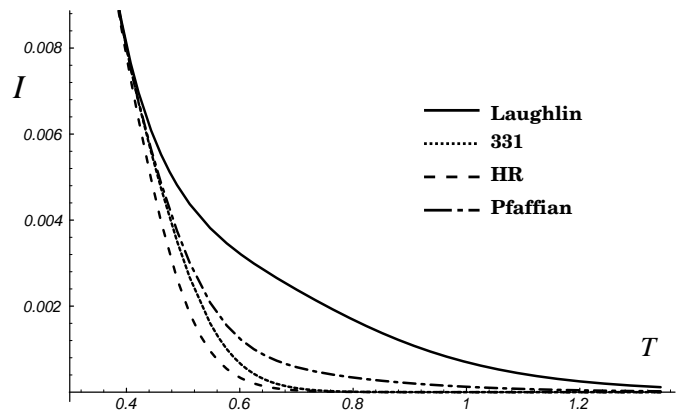}
\caption[teperaturenu1_2]
{Temperature dependence of persistent currents at $\phi=-0.05$.}
\label{t2}
\end{figure}

Fig.{\ref{t2}} shows the temperature dependence of 
the persistent currents near the origin,  $\phi=-0.05$.  
The currents decay exponentially as the temperature is highered. 
The magnitude of the currents for  paired states 
decays faster than the one for the Laughlin state.  
The temperature dependence 
is also different among paired states. 

Anomalous oscillations of the persistent currents 
are  explained  from the BCS pairing of electrons in the bulk and edge 
 of paired states.  
Naively, the edge persistent currents 
which we have calculated should have a period $\phi_0/2$  
 since the  bulk states of paired states are in a BCS 
superconducting phase and the order parameter has a charge $2e$. 
However as  there is no spontaneous
symmetry breaking of continuous symmetry in $1+1$ dimensions, 
 the edge states can not be a BCS condensate except 
at zero temperature. 
Then, from the behavior of the currents we have found, we see 
that  as we lower the temperature, the edge states become
 closer to a BCS condensate,  but  the phase transition of the edge
 states does not occur at finite temperatures. 
Thus the BCS pairing structure in the edge states  becomes 
 stronger   at lower temperatures,  but the condensation 
 only occurs at zero temperature.
        
This phenomenon may be seen as   an interesting  bridge 
between superconductivity in 2+1 dimensions
 and superconductivity 1+1 dimensions.   

The suppression of the persistent current means that  the 
increase of the viscosity of paired quantum Hall fluids 
from the hydrodynamical point 
of view  which is often used to describe the edge excitations 
for the fractional quantum Hall droplet \cite{wen}. 
It implies that the presence of strong pairing of  electrons 
increases the viscosity of the fluids  beyond the Laughlin state.

Thus we  see that the predicted behaviors 
 of  persistent edge currents can be used as   
a  method to distinguish the bulk topological order.  
Especially the flux dependence 
can reveal the pairing of electrons in the bulk of 
fractional quantum Hall states. 

Experimentally, the magnitude of the persistent currents is 
 preferably measurable at low temperatures and small samples.   
As we have predicted, experiments at the even-denominator 
plateau, or  in the double layers
may detect the anomalous oscillations of the persistent current.

{\it Acknowledgement.} The author would like to thank 
T.Ando, D.Lidsky, M.Kohmoto, J.Shiraishi and 
for discussions and M.R.Geller  and  especially 
M.Flohr for useful correspondence.

\vskip 0.6in

\def\NP{{ Nucl. Phys.\ }}
\def\PRL{{ Phys. Rev. Lett.\ }}
\def\PL{{ Phys. Lett.\ }}
\def\PR{{ Phys. Rev.\ }}
\def\CMP{{ Comm. Math. Phys.\ }}
\def\IJMP{{ Int. J. Mod. Phys.\ }}
\def\MPL{{ Mod. Phys. Lett.\ }}
\def\RMP{{ Rev. Mod. Phys.\ }}
\def\AP{{ Ann. Phys. (NY)\ }}

\end{multicols}

\end{document}